\documentclass[twocolumn]{aastex701}

\usepackage{amsmath}
\graphicspath{{./}{figures/}}

\newcommand{\tic}{TIC~402710280}
\newcommand{\planet}{TIC~402710280.01}
\newcommand{\tess}{\textit{TESS}}
\newcommand{\gaia}{\textit{Gaia}}
\newcommand{\rearth}{R_\oplus}
\newcommand{\rsun}{R_\odot}
\newcommand{\msun}{M_\odot}

\shorttitle{Sub-Neptune Candidate TIC 402710280.01}
\shortauthors{Vora}

\begin{document}

\title{A 36.34-day Sub-Neptune Candidate Transiting the Nearby K Dwarf TIC 402710280}

\author[orcid=0009-0003-8778-9883,gname=Rishit,sname=Vora]{Rishit Vora}
\affiliation{Independent Researcher, Bengaluru, India}
\email[show]{rishitv@protonmail.com}

\begin{abstract}
Transiting sub-Neptunes on orbits longer than a month are rare around bright stars. We report a candidate with a 36.34-day period around \tic, a $V = 10.0$ K dwarf at 66.8~pc that the Transiting Exoplanet Survey Satellite (\tess) has observed only in full-frame images. Four transits from 2021 to 2026, reduced by three independent pipelines, have consistent depths (mean $2092 \pm 47$~ppm) and a duration of $5.27 \pm 0.05$~h. Two are 36.36 days apart. The light curves exclude every shorter period consistent with both down to one day, and the duration disfavors still shorter ones, leaving $P = 36.33986 \pm 0.00051$~d. The candidate's radius is $3.75 \pm 0.22\,\rearth$ and its equilibrium temperature $498 \pm 21$~K. Difference-image centroids place the signal $1.5'' \pm 1.3''$ from the target, and eclipse models of the only nearby \gaia\ source, $3.5''$ away, fit worse than a planet by $\Delta\chi^2 \geq 142$ for every main-sequence pair. \texttt{TRICERATOPS} gives a false-positive probability of $0.003 \pm 0.004$; without high-resolution imaging, the object remains a candidate. The transit times scatter about a linear ephemeris by 21.8~min (rms), and an independent reduction reproduces the largest departure. Four transits cannot separate timing systematics from transit-timing variations. A second signal, of $407 \pm 26$~ppm at 3.78~days, may come from the target, an unresolved companion, or the background star. We predict a transit in \tess\ Sector~114 on 2027 February~19.
\end{abstract}


\section{Introduction} \label{sec:intro}

The NASA Exoplanet Archive lists six confirmed transiting planets of 3--4\,$\rearth$ around K dwarfs ($T_{\rm eff} = 3900$--5300~K) brighter than $V = 10.5$, and only one of them, TOI-2076~d \citep[$P = 35.1$~d;][]{Hedges2021}, has a period longer than 30~days.\footnote{Planetary Systems Composite Parameters table \citep{NEAcomp}, queried on 2026 October~21 for transiting planets with $3 \leq R_p/\rearth \leq 4$, $V < 10.5$, and $3900 \leq T_{\rm eff}/{\rm K} \leq 5300$; the query and the resulting table are in the data record \citep{DataRecord}.} Such planets are warm rather than strongly irradiated, and around stars this bright their masses are measurable with high-precision radial velocities. A single sector of the Transiting Exoplanet Survey Satellite \citep[\tess;][]{Ricker2015}, about 27~d long, records at most one transit of such a planet, so finding them requires combining sectors observed years apart.

Stars observed only in full-frame images lie outside the two-minute transit search of the Science Processing Operations Center \citep[SPOC;][]{Jenkins2016}. Their long-period planets must be found in full-frame light curves, such as those of the Quick-Look Pipeline \citep[QLP;][]{Huang2020a, Huang2020b} and TESS-SPOC \citep{Caldwell2020}, which reach the most recent images only weeks to months after they are taken. \tic\ shows the difficulty. Until 2026, its public light curves held two transits four years apart, and the only intervening epoch that \tess\ observed fell on a scattered-light ramp, so the period could not be measured from public data.

Here we report a sub-Neptune candidate with a period of 36.34~days around \tic, a K dwarf at 67~pc. We found its first two transits in public full-frame light curves and fixed the period with our own photometry of the 2026 full-frame images. As of 2026 October~18, the star is neither a \tess\ Object of Interest \citep[TOI;][]{Guerrero2021} nor a community TOI, SPOC has reported no threshold-crossing event for it, and neither the Exoplanet Follow-up Observing Program (ExoFOP), the NASA Exoplanet Archive \citep{Akeson2013}, nor the NASA Astrophysics Data System lists planet parameters, follow-up observations, or publications for it. We call the candidate \planet, the designation ExoFOP gives the first community TOI on a star, and will submit it as a community TOI once this paper is published in a refereed journal, as ExoFOP requires.

\section{Host Star} \label{sec:star}

Table~\ref{tab:star} lists the properties of \tic. We adopt the stellar parameters of the \tess\ Input Catalog v8.2 \citep[TIC;][]{Stassun2019}. The \gaia\ DR3 \citep{Gaia2016, GaiaDR3} astrophysical parameters agree with them: the GSP-Phot effective temperature is 5089~K \citep{Andrae2023}, and the FLAME radius and mass, $0.843 \pm 0.018\,\rsun$ and $0.81 \pm 0.04\,\msun$ \citep{Creevey2023}, lie within $1\sigma$ of the TIC values. The FLAME radius would make the candidate 3\% larger.

\gaia\ DR3 reports a renormalized unit weight error \citep[RUWE;][]{Lindegren2021} of $1.00$, no multi-peaked image windows (\texttt{ipd\_frac\_multi\_peak} $= 0$), and no non-single-star solution, as expected for a single star. These indicators do not exclude a faint companion at small separation.

The light curves show no rotational modulation. The strongest Lomb--Scargle peaks of individual sectors lie between 5 and 13~d, differ from sector to sector, and stay below 0.1\% peak to peak in the TESS-SPOC light curves, whose systematics correction suppresses longer periods. We find no flares: no light curve has two or more consecutive cadences more than $4\sigma$ above its running median, except during one scattered-light episode in Sector~93 (BTJD~3830.6), when the excess follows the background. There is therefore no stellar signal against which to phase the transits.

\begin{deluxetable}{lcc}
\tablecaption{Properties of \tic \label{tab:star}}
\tablehead{\colhead{Parameter} & \colhead{Value} & \colhead{Source}}
\startdata
R.A. (J2000) & 16:34:05.67 & TIC \\
Decl. (J2000) & $-$63:32:58.8 & TIC \\
\gaia\ DR3 ID & 5828441289812582400 & \gaia \\
$\mu_\alpha \cos\delta$ (mas yr$^{-1}$) & $59.909$ & \gaia \\
$\mu_\delta$ (mas yr$^{-1}$) & $13.685$ & \gaia \\
Parallax (mas) & $14.964 \pm 0.014$ & \gaia \\
Distance (pc) & $66.83 \pm 0.06$ & \gaia \\
RUWE & $1.00$ & \gaia \\
$T$ (mag) & $9.243 \pm 0.006$ & TIC \\
$G$ (mag) & $9.781$ & \gaia \\
$G_{\rm BP} - G_{\rm RP}$ (mag) & $1.023$ & \gaia \\
$V$ (mag) & $9.982 \pm 0.004$ & TIC \\
$J$ (mag) & $8.528 \pm 0.026$ & TIC \\
$K_s$ (mag) & $8.026$ & TIC \\
Contamination ratio & $0.060$ & TIC \\
$T_{\rm eff}$ (K) & $5182 \pm 121$ & TIC \\
$\log g$ (cgs) & $4.56 \pm 0.08$ & TIC \\
$R_\star$ ($\rsun$) & $0.817 \pm 0.047$ & TIC \\
$M_\star$ ($\msun$) & $0.88 \pm 0.11$ & TIC \\
$\rho_\star$ ($\rho_\odot$) & $1.61 \pm 0.38$ & derived \\
$L_\star$ ($L_\odot$) & $0.430 \pm 0.005$ & \gaia\ FLAME \\
\enddata
\tablecomments{$\rho_\star$ is computed from the TIC mass and radius. The contamination ratio is the TIC estimate of the flux from other stars relative to the target's within a \tess\ aperture. $J$ and $K_s$ are 2MASS magnitudes.}
\end{deluxetable}

\section{Observations and Photometry} \label{sec:data}

\tess\ observed \tic\ only in full-frame images, in Sectors~12 (2019), 39 (2021), 66 (2023), 93 (2025), and 100--103 (2026), at cadences of 1800~s in the prime mission, 600~s in the first extended mission, and 200~s thereafter. We use the TESS-SPOC light curves \citep[pre-search data conditioning flux;][]{Caldwell2020, TESSSPOC} of Sectors~12, 39, and~66 and the QLP light curve \citep[systematics-corrected flux;][]{Huang2020a, QLPdata} of Sector~93. For Sectors~100--103 we extracted light curves from $21 \times 21$-pixel cutouts \citep{Brasseur2019} of the SPOC-calibrated full-frame images \citep{TESSFFI}. The background of each cadence is the median of the faintest 40\% of pixels more than 4~pixels from the target. The aperture is a circle whose radius, chosen between 1 and 4~pixels in steps of 0.5, minimizes the one-hour scatter of the detrended light curve; the adopted radii are 1.5--2.5~pixels. We removed cadences flagged by SPOC and scattered-light glints, in which the background exceeds its one-day running median by more than eight times its median absolute deviation. QLP light curves of Sectors~100--103 were released at MAST between 2026 March and July but were absent from the public cloud copy of the archive that our search used. We use them only to check our transit times and depths (\S\ref{sec:ttv}).

The time stamps of the calibrated full-frame images are barycentric for a reference position on the detector, not for the target. Recomputing them for the target's position changes them by $-48$~s and $-20$~s at the Sector~100 and 102 transits. The TESS-SPOC and QLP light curves we fit already refer to the target: their time stamps precede those of the same cadences in the calibrated images by 20.7~s, in agreement with the 20--23~s we compute.

The TESS-SPOC fluxes are corrected for light from neighboring stars (\texttt{CROWDSAP} $= 0.92$--$0.94$); our photometry is not. The depths from the two kinds of light curve agree within their 4--7\% uncertainties (\S\ref{sec:times}), so any difference in dilution between them is smaller than the measurement precision.

For the transit analysis, we extract each transit in a window of $\pm 0.8$~d and normalize it by a quadratic function of time fitted to the out-of-transit data.

\section{Detection and Period} \label{sec:period}

We found the candidate in a search of 10,324,827 public light curves for individual transit-like events: QLP light curves of Sectors~1--98 and SPOC two-minute light curves \citep{TESSLC} of Sectors~99--107. We applied box filters of ten durations between 0.75 and 18~h and kept events above $6\sigma$ relative to the scatter of each filtered series. After we removed systematics shared by many stars, events at data edges and gaps, and non-planetary shapes, 2263 stars retained single transit-like events. We vetted the 142 of them brighter than $T = 10.5$ individually, at the pixel level and against catalogs of known objects, and searched the 27 that survived without a known period for further transits in our photometry of Sectors~99--108.

\begin{figure*}
\epsscale{1.15}
\plotone{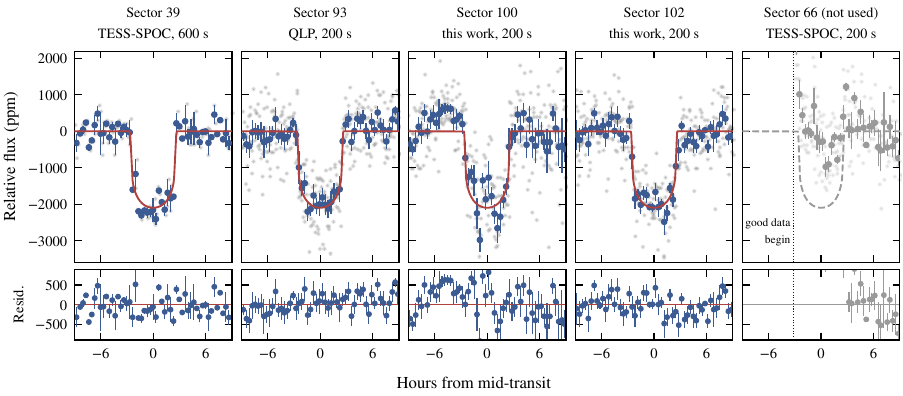}
\caption{Transits of \planet. Gray points are individual cadences, and blue points are 20-minute bins with error bars equal to the standard error of the mean. Each window is centered on its measured mid-time (Table~\ref{tab:times}), and the red curve is the median transit model of \S\ref{sec:fit}; the lower panels show the binned residuals. Our Sector~100 and 102 photometry has a larger per-cadence scatter than the QLP light curve at the same cadence (Table~\ref{tab:times}) because our apertures are smaller and are not optimized per cadence. The right-hand panel shows the Sector~66 window at the time predicted by the linear ephemeris; cadences without quality flags begin at the dotted line, and the window enters no fit (\S\ref{sec:period}).\label{fig:transits}}
\end{figure*}

The public light curves of \tic\ contain two dips of matching depth and duration, in Sectors~39 and 93, 1453.58~d apart. Our photometry revealed two more transits, in Sectors~100 and 102, 36.36~d apart (Figure~\ref{fig:transits}). Two transits 36.36~d apart allow only periods of $36.36/k$~d for integer $k$. We tested every $k$ from 2 to 36, down to 1.0~d, by measuring the depth at every transit time each period implies within the data, excluding the four observed transits. The exclusion significances below include the uncertainty of the observed depth. They use white-noise uncertainties for the windows, and correlated noise as large as that of the transit windows ($\beta \leq 1.5$, Table~\ref{tab:times}) would reduce them by up to a third. At $P/2$, transits are expected in three covered windows (Sectors~39, 101, and~102) and are absent from each at $\geq 23.8\sigma$; the other $P/2$ epochs in the observed sectors fall in the mid-sector gap of Sector~12, before the first good cadence of Sector~100, or after the end of the Sector~93 data. At $P/3$, transits are absent from eight windows at $\geq 23.5\sigma$, and at $P/4$ and $P/5$ from 27 more at $\geq 17.1\sigma$. Each shorter period, from $P/6$ to $P/36$, places 19 to 152 transits in the data, and their mean depth falls short of the observed one by at least $41\sigma$. Periods below 1.0~d are strongly disfavored by the duration. The longest transit such an orbit allows, centered at apastron on an eccentric orbit whose periastron touches the stellar surface, lasts 4.8~h around a star of the cataloged density, compared with the observed $5.27 \pm 0.05$~h, and reaches the observed duration only for a density 1.9$\sigma$ below the cataloged value. At the adopted period, the Sector~12 data (BTJD 1624.9--1652.9, where BTJD $=$ BJD$_{\rm TDB} - 2457000$) fall between transits at BTJD~1621.5 and 1657.8, Sector~101 contains no transit, and the Sector~103 transit, at BTJD~4165.26, falls in the data gap between BTJD~4164.44 and 4166.79. A period of 36.34~d is therefore the only one consistent with every covered epoch.

\begin{figure}
\plotone{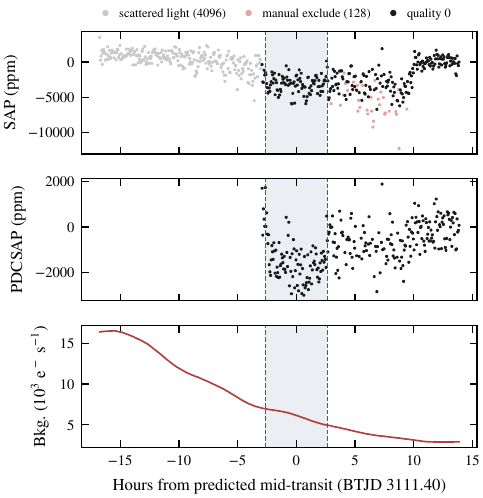}
\caption{TESS-SPOC data of Sector~66 around the predicted transit, whose duration is shaded and whose ingress and egress are marked by dashed lines. From top: simple-aperture flux, with cadences flagged by SPOC for scattered light (quality value 4096) in gray and as manually excluded (value 128) in pink; pre-search data conditioning flux, quality 0 only; and the background.\label{fig:s66}}
\end{figure}

The transit expected in Sector~66 (BTJD~3111.40) coincides with strong scattered light (Figure~\ref{fig:s66}). The background falls from $1.5 \times 10^4$ to $2.9 \times 10^3$~e$^-$~s$^{-1}$ across the window, the first cadences without quality flags begin 3.0~h before the predicted mid-time (0.4~h before ingress), and QLP flags every cadence within 12~h of it. The transit is covered, but the baseline before it is not. Fitted with the shape of \S\ref{sec:fit} and a free mid-time, the event has depths of 1380, 1270, and 1620~ppm (uncertainties 100--160~ppm) for constant, linear, and quadratic baselines and lies $-5$ to $+2$~min from the linear ephemeris of \S\ref{sec:ttv}. Its timing is robust to the baseline, but its depth is not, so we exclude Sector~66 from all fits of the transits and from the ephemeris.

\section{Transit Analysis} \label{sec:analysis}

\subsection{Individual Transit Times and Depths} \label{sec:times}

We timed each transit by fixing the shape to the median model of \S\ref{sec:fit} and scanning the mid-time on a 0.5-minute grid over $\pm 2$~h, refitting a quadratic baseline across the $\pm 0.8$~d window at each step. The $1\sigma$ uncertainty is the half-width of the $\Delta\chi^2 = 1$ interval multiplied by the red-noise factor $\beta$ \citep{Pont2006, Winn2008}, the median ratio of the binned residual scatter (10--60-minute bins) to its white-noise expectation. We measured each depth with the shape fixed and the amplitude free. Table~\ref{tab:times} lists the results.

The four depths agree ($\chi^2 = 2.5$ for 3 degrees of freedom), although they come from three independent reductions with different apertures, cadences, and crowding corrections. This agreement argues against an origin in any one pipeline and against aperture-dependent dilution. Their weighted mean is $2092 \pm 47$~ppm.

\begin{deluxetable*}{lcccccccc}
\tablecaption{Individual transits of \planet \label{tab:times}}
\tablehead{\colhead{Sector} & \colhead{Light curve} & \colhead{Cadence (s)} & \colhead{$\sigma$ (ppm)} & \colhead{$\beta$} & \colhead{Epoch} & \colhead{$T_c$ (BJD$_{\rm TDB}$)} & \colhead{$O - C$ (min)} & \colhead{Depth (ppm)}}
\startdata
39 & TESS-SPOC & 600 & 355 & 1.00 & 0 & $2459384.6094 \pm 0.0013$ & $+4.4$ & $2203 \pm 86$ \\
93 & QLP & 200 & 594 & 1.02 & 40 & $2460838.1847 \pm 0.0012$ & $-23.2$ & $2025 \pm 84$ \\
100 & This work & 200 & 702 & 1.47 & 47 & $2461092.5845 \pm 0.0023$ & $+6.8$ & $2078 \pm 144$ \\
102 & This work & 200 & 608 & 1.08 & 48 & $2461128.9446 \pm 0.0017$ & $+35.9$ & $2053 \pm 91$ \\
\enddata
\tablecomments{$\sigma$ is the out-of-transit scatter per cadence and $\beta$ the red-noise factor of each window; every window spans $\pm 0.8$~d and has a quadratic baseline. $O-C$ is relative to the linear ephemeris of Table~\ref{tab:params}. Depths are the model depth at mid-transit for a free amplitude with the shape fixed.}
\end{deluxetable*}

\subsection{Transit Shape} \label{sec:fit}

We fitted the four transits jointly with \texttt{batman} \citep{Kreidberg2015}, integrating over each light curve's exposure time and centering each window on its measured mid-time so that the timing scatter (\S\ref{sec:ttv}) does not smear the stacked profile. The free parameters are $R_p/R_\star$, $b$, and $\log_{10}\rho_\star$, with uniform priors; $a/R_\star$ follows from $\rho_\star$ and $P$ for a circular orbit \citep{Seager2003}. The quadratic limb-darkening coefficients are fixed at $(u_1, u_2) = (0.39, 0.21)$, interpolated in $T_{\rm eff}$ from the \tess-band tables of \citet{Claret2017} at $\log g = 4.5$. Each window's flux uncertainty is its out-of-transit point-to-point scatter multiplied by its $\beta$. We sampled the posterior with \texttt{emcee} \citep{Goodman2010, ForemanMackey2013}, running 40 walkers for 30,000 steps and discarding the first third; the chain spans more than 90 integrated autocorrelation times for every parameter. Table~\ref{tab:params} gives the medians and 68\% intervals. The best-fitting model has a reduced $\chi^2$ of 0.84 for 2227 data points, so the $\beta$-scaled uncertainties are slightly conservative.

The impact parameter is below 0.54 at 95\% confidence. The fitted stellar density, $1.57^{+0.12}_{-0.31}\,\rho_\odot$, agrees with the $1.61 \pm 0.38\,\rho_\odot$ derived from the TIC mass and radius, as expected for a circular orbit around a star of the cataloged density. Fitting the limb darkening as well, with the uninformative parametrization of \citet{Kipping2013}, changes $R_p/R_\star$ to $0.0426 \pm 0.0006$ and $b$ to $0.21^{+0.19}_{-0.15}$, both within $1\sigma$ of the adopted values.

\subsection{Ephemeris and Timing Residuals} \label{sec:ttv}

\begin{figure}
\plotone{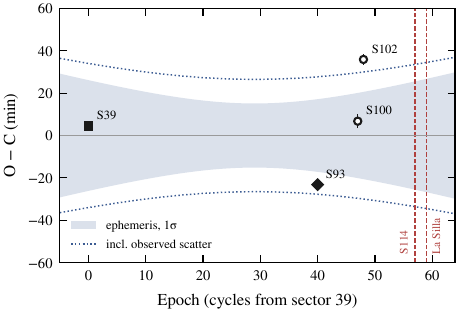}
\caption{Transit-timing residuals relative to the linear ephemeris. Symbols denote the light-curve source: TESS-SPOC (square), QLP (diamond), and our photometry (open circles). The formal uncertainties, 1.7--3.3~min, are smaller than the symbols. The shaded band is the $1\sigma$ uncertainty of the ephemeris, and the dotted curves add the 21.8-min rms scatter of the residuals in quadrature. Dashed lines mark epochs 57 (2027 February~19, \tess\ Sector~114) and 59 (2027 May~3, observable from La Silla).\label{fig:oc}}
\end{figure}

A weighted linear fit to the four mid-times gives $P = 36.33986 \pm 0.00051$~d and $T_0 = 2460838.2008 \pm 0.0119$~BJD$_{\rm TDB}$, with the reference epoch at the Sector~93 transit (epoch 40). The weighted mean epoch is 29, so $T_0$ and $P$ are correlated (correlation coefficient $+0.47$), and the predictions of \S\ref{sec:predictions} include their covariance. The residuals (Table~\ref{tab:times}, Figure~\ref{fig:oc}) are $+4.4$, $-23.2$, $+6.8$, and $+35.9$~min. Their rms of 21.8~min far exceeds the formal uncertainties of 1.7--3.3~min ($\chi^2 = 417$ for 2 degrees of freedom), and we scale the ephemeris uncertainties by $\sqrt{\chi^2/\nu}$. The largest single contribution is the interval between the consecutive Sector~100 and 102 transits, which exceeds $P$ by 29.2~min.

We looked for an instrumental cause. Changing the baseline (constant, linear, or quadratic) or the window half-width (0.6, 0.8, or 1.0~d) moves no mid-time by more than 2.3~min. The Sector~93 mid-time agrees within 1.9~min across every QLP flux series: the systematics-corrected flux, the raw flux, and the detrended fluxes of the optimal, small, and large apertures. That window is complete (688 of 691 cadences, no gap longer than 7~min) and shows no background or centroid feature at the transit. The time stamps of all four light curves refer to the target's barycentric position (\S\ref{sec:data}). Masking the transits of the 3.78-day signal (\S\ref{sec:second}), one of which falls 4~h before the Sector~93 transit, moves no mid-time by more than 0.1~min.

The QLP light curves of Sectors~100 and~102 provide an independent reduction of the same images, but their time stamps must first be corrected. Matched to the full-frame images by cadence number, they precede the mid-exposure times of the images, corrected to the target, by 121 and 139~s at the two transits; in Sector~93 the two agree within 3~s. The offset changes by $-6.3$ and $-6.1$~s per day, the negatives of the rates at which the target's barycentric correction changes (6.3 and 6.1~s per day), and the QLP times of Sectors~100 and~102 are linear in cadence number to within 0.01~s across each sector. Both results point to a single barycentric correction applied to every cadence of each sector. The offset vanishes, and the applied correction equals the target's, near BTJD~4072, two days before the first Sector~100 cadence (BTJD~4074.0), and near BTJD~4107, during Sector~101 and three weeks before the first Sector~102 cadence (BTJD~4127.0). QLP adopted the photometry of \citet{Han2023} from Sector~94 onward \citep{Petitpas2026}; the Sector~93 time stamps, which follow the barycentric correction, precede that change. We have reported the offset to the QLP team.

With this correction, and timed as in \S\ref{sec:times}, the QLP transits fall 4.1 and 4.6~min earlier than ours (0.7 and 1.5 times the combined uncertainty), and the interval between them exceeds $P$ by 28.7~min, compared with 29.2~min in our photometry. Our photometry of the Sector~39 and 93 transits, reduced in the same way, gives mid-times within 0.8~min of the TESS-SPOC and QLP values. The excess in the Sector~100--102 interval therefore does not arise in our reductions.

The QLP depths of the Sector~100 and 102 transits, $1757 \pm 170$ and $1666 \pm 83$~ppm, are lower than ours by 1.4 and 3.1 times the combined uncertainty, whereas the Sector~93 QLP depth, measured before the change of method, is $2025 \pm 84$~ppm and agrees with the others. Our photometry is not corrected for neighboring stars, which would make its depths too shallow rather than too deep, so the difference does not come from our treatment of crowding. We have not identified its cause, and neither QLP light curve enters our fits. Adopting the QLP depths of these two transits would lower the mean depth to $1943 \pm 47$~ppm and the radius to $3.61\,\rearth$, but the four depths would then disagree ($\chi^2 = 22$ for 3 degrees of freedom). Adopting the QLP mid-times would change $P$ by 4~s and the timing rms to 19.3~min.

We identify no instrumental cause of the timing scatter in the light curves we fit. Four epochs cannot distinguish timing systematics from transit-timing variations induced by another body, and further transits will show whether the residuals are periodic. The predictions of \S\ref{sec:predictions} include the observed scatter.

\begin{deluxetable}{lc}
\tablecaption{Parameters of \planet \label{tab:params}}
\tablehead{\colhead{Parameter} & \colhead{Value}}
\startdata
\multicolumn{2}{c}{\textit{Ephemeris}} \\
$P$ (d) & $36.33986 \pm 0.00051$ \\
$T_0$ (BJD$_{\rm TDB}$) & $2460838.2008 \pm 0.0119$ \\
Timing residual rms (min) & 21.8 \\
\multicolumn{2}{c}{\textit{Fitted}} \\
$R_p/R_\star$ & $0.04209^{+0.00067}_{-0.00051}$ \\
$b$ & $0.22^{+0.20}_{-0.15}$ \\
$\rho_\star$ ($\rho_\odot$) & $1.57^{+0.12}_{-0.31}$ \\
\multicolumn{2}{c}{\textit{Fixed}} \\
$e$ & 0 \\
$(u_1, u_2)$ & $(0.39, 0.21)$ \\
\multicolumn{2}{c}{\textit{Derived}} \\
Depth at mid-transit (ppm) & $2092 \pm 47$ \\
$T_{14}$ (h) & $5.27 \pm 0.05$ \\
$a/R_\star$ & $53.6^{+1.3}_{-3.7}$ \\
$R_p$ ($\rearth$) & $3.75 \pm 0.22$ \\
$a$ (AU) & $0.206 \pm 0.009$ \\
$S$ ($S_\oplus$) & $10.2 \pm 1.7$ \\
$T_{\rm eq}$ (K) & $498 \pm 21$ \\
\enddata
\tablecomments{Medians with 68\% intervals. The depth is the weighted mean of the four measurements of Table~\ref{tab:times}. $T_{\rm eq}$ assumes zero Bond albedo and full heat redistribution. $R_p$, $a$, $S$, and $T_{\rm eq}$ use the stellar parameters of Table~\ref{tab:star}.}
\end{deluxetable}

\section{Vetting} \label{sec:vetting}

Without high-resolution imaging or spectroscopy, the light curves and the \gaia\ catalog test four things: whether the signal comes from the target's position (\S\ref{sec:centroid}), whether the one cataloged star close enough to contribute can produce it (\S\ref{sec:blend}), whether the target is itself an eclipsing binary (\S\ref{sec:secondary}), and how probable the remaining false-positive scenarios are, given the field (\S\ref{sec:fpp}). They cannot exclude an unresolved companion within about $1''$.

\subsection{Location of the Signal} \label{sec:centroid}

\begin{figure*}
\epsscale{1.1}
\plotone{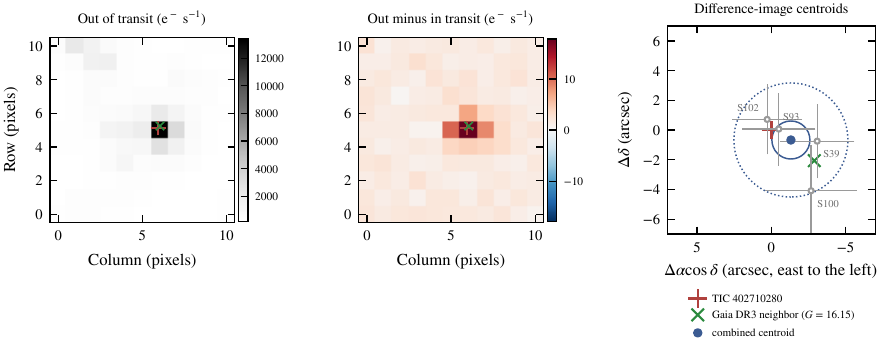}
\caption{Location of the signal. Left and center: the out-of-transit image and the difference image (out of transit minus in transit) of the Sector~39 transit, which has the highest signal-to-noise ratio. The plus marks \tic\ and the cross the background neighbor; the two are not separated at the 21$''$ pixel scale. Right: the centroid offsets of the four transits (gray) and their combination (blue), with $1\sigma$ (solid) and $3\sigma$ (dotted) circles, relative to \tic. The cross marks the neighbor at the mean epoch of the transits.\label{fig:diff}}
\end{figure*}

For each transit we formed a difference image (out of transit minus in transit) from the full-frame cutouts and fitted a Gaussian point-spread function to it \citep[cf.][]{Bryson2013, Twicken2018}. We measured each offset relative to the target's own fitted position in the out-of-transit image, which removes errors in the world-coordinate solution and the effect of proper motion. The four transits combined place the signal $1.5'' \pm 1.3''$ from \tic\ (Figure~\ref{fig:diff}). Applied to 40 confirmed \tess\ planets, the method placed 30 on their host star, left 5 unresolved from a nearby source, returned an offset without a cataloged counterpart for 3, failed for 2, and placed none on a neighboring star.

The centroid lies $1.2\sigma$ from the target and $1.6\sigma$ from one neighbor. Every other \gaia\ DR3 source bright enough to produce the observed depth lies $21\sigma$ or more away and is excluded; this one is not. The neighbor, \gaia\ DR3 5828441289812582656 ($G = 16.15$, $\Delta G = 6.37$), has a parallax of $0.69 \pm 0.08$~mas and is therefore in the background. Because of the target's proper motion, the separation grows from $3.10''$ at the \gaia\ epoch (2016.0) to $3.4''$--$3.6''$ during the transits.

\subsection{Blended Eclipsing Binary} \label{sec:blend}

\begin{figure}
\plotone{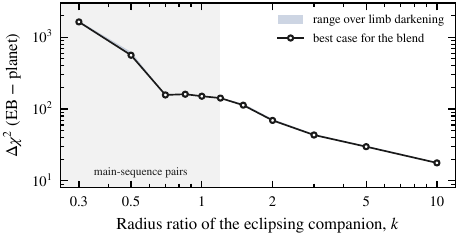}
\caption{Diluted-eclipse models for the background neighbor, compared with the planet model. For each companion radius ratio $k$, the curve gives the smallest $\chi^2$ difference over impact parameter, $a/R_\star$, dilution, and limb darkening, and the band gives the range over the three limb-darkening laws. The shaded region, $k \leq 1.2$, contains every main-sequence pair.\label{fig:blend}}
\end{figure}

Diluted by \tic, the neighbor could produce the observed depth only by losing about 74\% of its light in eclipse. We fitted the four transits with eclipse models of the neighbor for companion-to-star radius ratios $k$ from 0.3 to 10, with the impact parameter and $a/R_\star$ free, the dilution free but bounded at 0.4\% of the total flux (40\% above the star's share, to allow for its unknown \tess-band color), and three limb-darkening laws (Figure~\ref{fig:blend}). For $k \leq 1.2$, every eclipse model fits worse than the planet model by $\Delta\chi^2 \geq 142$ for 2227 data points. Larger companions fit better ($\Delta\chi^2 = 113$, 70, 44, 30, and 18 for $k = 1.5$, 2, 3, 5, and 10), because a total eclipse with a short ingress resembles a transit, but only if the eclipsed star supplies about 70\% of the neighbor's light and a companion 1.5 to 10 times its radius supplies the rest. The companion would then need a surface brightness 5 ($k = 1.5$) to 200 ($k = 10$) times lower than the eclipsed star's, which no main-sequence pair has; only a hot, compact eclipsed star could produce such an eclipse. A blend with the neighbor is therefore disfavored, and a ground-based light curve with an aperture that excludes the neighbor would settle the question.

\subsection{Secondary Eclipse and Odd--Even Depths} \label{sec:secondary}

The light curve shows no eclipse at orbital phase 0.5, which Sectors~39, 101, and~102 sample: the weighted depth there is $-39 \pm 39$~ppm, below 77~ppm at $3\sigma$. A search over phases 0.3 to 0.7, which covers secondary eclipses of eccentric orbits with $|e\cos\omega| \lesssim 0.3$, finds no consistent signal. The $3\sigma$ upper limit is below 160~ppm at half of these phases and below 365~ppm at all of them. The most significant box, at phase 0.449 ($204 \pm 53$~ppm), is dominated by data from the last 0.5~d of Sector~93 and is not reproduced in Sectors~39 and 101. These limits exclude an eclipsing companion with a surface brightness comparable to the target's. Only one transit (Sector~100) falls on an odd epoch, so odd and even depths cannot be compared usefully, but the agreement of all four depths (\S\ref{sec:times}) disfavors an eclipsing binary at twice the period with unequal eclipses.

\subsection{Statistical False-Positive Probability} \label{sec:fpp}

Statistical validation with \texttt{TRICERATOPS} \citep{Giacalone2020, Giacalone2021} requires high-resolution imaging, and none exists for this star. Without a contrast curve, unresolved companions are constrained only by population priors, so the probabilities below do not validate the candidate. We computed scenario probabilities from the four transits stacked on their measured mid-times (10-minute bins), a depth of 2100~ppm, the photometric apertures of Sectors~39, 100, and~102, a circle of 2~pixels radius in Sector~93 (the QLP aperture has a radius of 3~pixels), and the \gaia\ DR3 field. Ten independent runs give a false-positive probability ${\rm FPP} = 0.003 \pm 0.004$ (range 0.000--0.012) and a nearby false-positive probability ${\rm NFPP} < 10^{-6}$.\footnote{With the photometric apertures, NFPP is numerically zero ($<10^{-20}$); with the default $5 \times 5$-pixel apertures of \texttt{TRICERATOPS}, ${\rm FPP} = 0.001$ and ${\rm NFPP} = 6 \times 10^{-7}$.} \texttt{TRICERATOPS} places the stars at their TIC positions, at epoch J2000, where the neighbor lies $2.3''$ from the target rather than the $3.4''$--$3.6''$ of the transits (\S\ref{sec:centroid}). The smaller separation puts more of the neighbor's light in the aperture, so the computed NFPP overstates the nearby false-positive probability. A planet on the target receives 85\% of the probability and a planet diluted by an unresolved bound companion 14\%; the leading false-positive scenario, a planet orbiting an unresolved companion, receives 0.3\%. These values meet the numerical criteria of \citet{Giacalone2021}, ${\rm FPP} < 0.015$ and ${\rm NFPP} < 0.001$, but without imaging we classify \planet\ as a planet candidate.

\section{A Second Periodic Signal} \label{sec:second}

\begin{figure}
\plotone{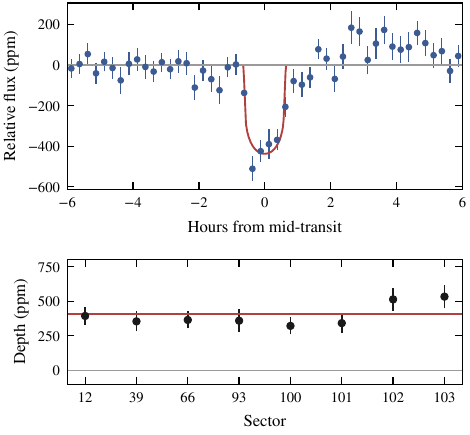}
\caption{The 3.78-day signal. Top: all eight sectors folded on its ephemeris, with the transits of \planet\ masked, in 15-minute bins (error bars are standard errors of the mean), and the median transit model. Bottom: the box depth in each sector, with uncertainties from boxes placed at 400 random phases; the line is the mean over all sectors.\label{fig:second}}
\end{figure}

With the transits of \planet\ masked, a box-least-squares search \citep{Kovacs2002} of the light curve finds a second periodic signal, with $P = 3.78371$~d, a depth of $407 \pm 26$~ppm, and a duration of about 1.3~h (Figure~\ref{fig:second}). The search detects it independently in Sectors~12--93, as the strongest periodicity (signal detection efficiency 14.4), and in Sectors~100--103, as the second strongest after its 5:1 alias at 18.92~d. The two ephemerides agree in phase across 650 cycles. The depths in the eight sectors agree with one another ($\chi^2 = 7.3$ for 7 degrees of freedom), the odd and even transits agree ($382 \pm 39$ and $437 \pm 46$~ppm), and no eclipse appears at phase 0.5 ($-61 \pm 26$~ppm).

A transit model with free stellar density gives $R_p/R_\star = 0.0196^{+0.0015}_{-0.0008}$, $b = 0.49 \pm 0.32$, and $\rho_\star = 8.3^{+3.7}_{-5.5}\,\rho_\odot$. On the target, the signal would correspond to a planet of $1.75 \pm 0.11\,\rearth$ on an inner orbit. The fitted density, $8.3^{+3.7}_{-5.5}\,\rho_\odot$, exceeds the target's $1.61 \pm 0.38\,\rho_\odot$. Only 13\% of the posterior lies below $2.37\,\rho_\odot$ (the target's value plus $2\sigma$), almost all of it at $b > 0.82$, so a transit of the target matches the short duration only if it is nearly grazing. The difference-image centroid of this weaker signal, $5.2'' \pm 6.7''$ from the target, cannot separate the target from the background neighbor, which would have to lose about 14\% of its light in eclipse to produce the signal. \texttt{TRICERATOPS}, applied as in \S\ref{sec:fpp} but with the 3-pixel QLP aperture in Sector~93, assigns 98\% of the probability to a planet around the target or an unresolved companion but gives ${\rm FPP} = 0.12 \pm 0.01$ and ${\rm NFPP} = 0.015 \pm 0.001$, above the NFPP limit of 0.001 below which \citet{Giacalone2021} consider nearby false positives excluded. We therefore do not claim this signal as a planet candidate: its source may be the target, an unresolved companion, or the neighbor. A ground-based light curve that resolves the neighbor would show the 14\% eclipse if the neighbor is the source.

\section{Discussion} \label{sec:discussion}

With $R_p = 3.75 \pm 0.22\,\rearth$, \planet\ lies above the radius valley \citep{Fulton2017}, among sub-Neptunes that probably retain a substantial volatile envelope. At $0.206$~AU it receives about ten times Earth's insolation. The mass--radius relation of \citet{ChenKipping2017} predicts $M_p \approx 14\,M_\oplus$, which would induce a radial-velocity semi-amplitude of about 2.8~m~s$^{-1}$, measurable with current high-precision spectrographs for a $V = 10.0$ star. For that mass, the transmission spectroscopy metric of \citet{Kempton2018} is about 70, below the threshold of 90 that \citet{Kempton2018} recommend for sub-Neptunes, so the planet is a moderate rather than a prime target for transmission spectroscopy. If confirmed, it would be the second planet of 3--4\,$\rearth$ with a period longer than 30~days around a K dwarf brighter than $V = 10.5$, after TOI-2076~d.

Three observations would test the candidate most efficiently: high-resolution imaging, to exclude unresolved companions and complete the statistical validation; seeing-limited photometry of a transit with an aperture that excludes the $3.5''$ neighbor, to locate both signals; and radial velocities, to measure the mass. Each could overturn the present interpretation. A companion within $1''$ bright enough to dilute the transit would change the radius and the false-positive probability, a 74\% eclipse of the neighbor at the time of a transit would identify the signal as a blend, and a semi-amplitude far above the few meters per second expected for a sub-Neptune would reveal a stellar or substellar companion.

\section{Predicted Transits} \label{sec:predictions}

Table~\ref{tab:predictions} lists three of the next transits predicted by the ephemeris of Table~\ref{tab:params}; the data record lists the next 30 \citep{DataRecord}. The quoted uncertainty, $\sigma_{\rm pred} = (\sigma_{T_0}^2 + n^2\sigma_P^2 + 2n\,{\rm Cov}(T_0, P) + \sigma_{\rm rms}^2)^{1/2}$ for $n$ cycles from the reference epoch, adds the observed residual scatter, $\sigma_{\rm rms} = 21.8$~min, to the ephemeris uncertainty, so that it describes an individual transit rather than the mean ephemeris. \tess\ is expected to observe \tic\ in Sectors~113 and~114 \citep{Burke2020}, and the transit of 2027 February~19 falls in Sector~114. The transit of 2027 May~3 is observable in full from La Silla. The test of these predictions, with the data to be used and the rules that decide its outcome, was registered before the data existed \citep{PreReg}. The registered $\sigma_{\rm pred}$ omits the covariance term (30.4 instead of 33.6~min at epoch 57), so the registered search over $\pm 3\sigma_{\rm pred}$ spans $\pm 2.7\sigma$ of the uncertainty given here; we keep the registered test unchanged. Under the registered rules, a transit absent from its window in data that cover it falsifies the period. A timing departure larger than the half-width of that window, four times the observed rms, would have the same signature.

\begin{deluxetable*}{ccccl}
\tablecaption{Predicted transits of \planet \label{tab:predictions}}
\tablehead{\colhead{Epoch} & \colhead{$T_c$ (BJD$_{\rm TDB}$)} & \colhead{$\sigma_{\rm pred}$ (min)} & \colhead{UTC at La Silla} & \colhead{Note}}
\startdata
57 & 2461455.9785 & 34 & 2027 Feb. 19, 11:29 & \tess\ Sector~114 \\
58 & 2461492.3183 & 34 & 2027 Mar. 27, 19:35 & daytime at La Silla \\
59 & 2461528.6582 & 34 & 2027 May 3, 03:41 & full transit at night at La Silla \\
\enddata
\tablecomments{Epochs count cycles from the Sector~39 transit. UTC times are topocentric at La Silla Observatory: the light-travel time and TDB$-$UTC are removed.}
\end{deluxetable*}

\begin{acknowledgments}
The author thanks Keshav Aggarwal (student visitor, Space Physics Laboratory, Vikram Sarabhai Space Centre, India) for an independent check of the archival transits, which prompted the comparison with the QLP light curves, and Pascal S.~D'Andrea (Institute for Particle Physics and Astrophysics, ETH Zurich, Switzerland) for an independent reanalysis of the four transits.

This paper includes data collected by the \tess\ mission, which are publicly available from the Mikulski Archive for Space Telescopes (MAST). Funding for the \tess\ mission is provided by NASA's Science Mission Directorate. This work has made use of data from the European Space Agency (ESA) mission \gaia\ (\url{https://www.cosmos.esa.int/gaia}), processed by the \gaia\ Data Processing and Analysis Consortium (DPAC, \url{https://www.cosmos.esa.int/web/gaia/dpac/consortium}). Funding for the DPAC has been provided by national institutions, in particular the institutions participating in the \gaia\ Multilateral Agreement. This research has made use of the Exoplanet Follow-up Observing Program (ExoFOP; DOI: 10.26134/ExoFOP5) website and of the NASA Exoplanet Archive, both operated by the California Institute of Technology under contract with the National Aeronautics and Space Administration under the Exoplanet Exploration Program.

The author used Claude (Anthropic), a large language model, throughout this work, under the author's direction and review: to write and debug the analysis code, to produce the figures from the analysis outputs, to search the literature, and to draft and edit the text. Every reference was checked against its DOI or ASCL record, and the analysis of the four transits can be reproduced from the archived code and data \citep{DataRecord}. Responsibility for the content lies entirely with the author.

\end{acknowledgments}

\facilities{TESS, Gaia}

\software{astropy \citep{Astropy2022}, astrocut \citep{Brasseur2019}, batman \citep{Kreidberg2015}, emcee \citep{ForemanMackey2013}, lightkurve \citep{Lightkurve2018}, matplotlib \citep{Hunter2007}, numpy \citep{Harris2020}, scipy \citep{Virtanen2020}, tess-point \citep{Burke2020}, TRICERATOPS \citep{Giacalone2020}}

\section*{Data Availability}
The light curves of all transit windows, our photometry of Sectors~100--103, the measured transit times, the comparison with the QLP light curves of Sectors~100 and~102 and the test of their time stamps, the alias and secondary-eclipse tests, the archive comparison sample, the posterior summaries and chains, the next 30 predicted transits, the analysis code, and the intermediate results from which the code fills in the manuscript are available on Zenodo \citep[version 4.0; all versions: doi:10.5281/zenodo.23390273]{DataRecord}. The search of \S\ref{sec:period}, which selected the star, belongs to a wider survey and is not part of the record. The pre-registered test of the 2027 transits is a separate Zenodo record \citep{PreReg}. All the \tess\ data used in this paper can be found in MAST: the TESS-SPOC light curves of Sectors~12, 39, and~66, \dataset[https://doi.org/10.17909/t9-wpz1-8s54]{https://doi.org/10.17909/t9-wpz1-8s54}; the QLP light curves of Sectors~1--98 (search) and Sectors~93, 100, and~102 (analysis), \dataset[https://doi.org/10.17909/t9-r086-e880]{https://doi.org/10.17909/t9-r086-e880}; the calibrated full-frame images of Sectors~39, 93, and 100--103, \dataset[https://doi.org/10.17909/0cp4-2j79]{https://doi.org/10.17909/0cp4-2j79}; and the SPOC two-minute light curves of Sectors~99--107 (search), \dataset[https://doi.org/10.17909/t9-nmc8-f686]{https://doi.org/10.17909/t9-nmc8-f686}. The stellar parameters are from the \tess\ Input Catalog, \dataset[https://doi.org/10.17909/fwdt-2x66]{https://doi.org/10.17909/fwdt-2x66}.

\bibliography{refs}{}
\bibliographystyle{aasjournalv7}

\end{document}